\documentclass[preprint,12pt]{elsarticle}
\journal{Journal of Computational Science}

\usepackage{graphicx}
\usepackage{color}

\usepackage{amssymb}
\usepackage{amsmath}
\usepackage{amsthm}
\usepackage{bm}

\usepackage{lscape}
\usepackage{subcaption}
\usepackage{hyperref}

\newcommand*\diff{\mathop{}\!\mathrm{d}}

\definecolor{forestgreen}{rgb}{0.13,0.54,0.13}

\begin{document}
\begin{frontmatter}

\title{A Lattice Boltzmann Method for Relativistic Rarefied Flows in $(2+1)$ Dimensions}

\author{L. Bazzanini$^{1}$, A. Gabbana$^{12}$, D. Simeoni$^{*134}$ S. Succi$^{56}$ and R. Tripiccione$^{1}$}
  
\address{ 
$^{1}$Universit\`a di Ferrara and INFN-Ferrara, I-44122 Ferrara,~Italy                                                                       \\
$^{2}$Eindhoven University of Technology, 5600 MB Eindhoven,~ The Netherlands                                                                 \\
$^{3}$Bergische Universit\"at Wuppertal, D-42119 Wuppertal,~Germany                                                                     \\
$^{4}$University of Cyprus, CY-1678 Nicosia,~Cyprus                         \\
$^{5}$Center for Life Nano Science @ La Sapienza, Italian Institute of Technology, Viale Regina Elena 295, I-00161 Roma,~Italy                                                                       \\
$^{6}$Istituto Applicazioni del Calcolo, National Research Council of Italy, Via dei Taurini 19, I-00185 Roma,~Italy                                                                       \\
}

\cortext[cor1]{Correspondence: d.simeoni@stimulate-ejd.eu}
 
\begin{abstract}
  We propose an extension to recently developed Relativistic Lattice Boltzmann
  solvers (RLBM),  which allows the simulation of flows close to the free
  streaming limit.  Following previous works
  [\href{https://doi.org/10.1103/PhysRevC.98.035201}{Phys. Rev. C 98 (2018)
  035201}], 
  we use product quadrature rules and select weights and nodes by  separately
  discretising the radial and the angular components. 

  This procedure facilitates the development of quadrature-based RLBM with
  increased isotropy levels, thus improving the accuracy of the method for  the
  simulation of flows beyond the hydrodynamic regime.

  In order to quantify the improvement of this discretisation procedure over
  existing methods,  we perform numerical tests of shock waves in one and two
  spatial dimensions  in various kinetic regimes across the hydrodynamic and
  the free-streaming limits.

\end{abstract}

\end{frontmatter}

%===================================================================================================
\section{Introduction}\label{sec:intro}
%===================================================================================================

Relativistic flows \cite{cattaneo-book-2011,lichnerowicz-book-1967,eckart-prl-1940,muller-zphys-1967,
landau-book-1987,israel-anp-1976,israel-prsl-1979} are of great relevance to
several research fields,  including astrophysics and cosmology
\cite{degroot-book-1980,rezzolla-book-2013} and 
high energy physics, in particular in connection with the study of the  quark
gluon plasma (QGP) \cite{florkowski-rpp-2018}. Relativistic hydrodynamics has
also found application in the context of condensed matter  physics, particularly
for the study of strongly correlated electronic fluids in exotic (mostly 2-d) 
materials, such as graphene sheets and Weyl semi-metals
\cite{lucas-jopcm-2018}.   

The mounting importance of the relativistic hydrodynamic approach for several
physics application areas  commands the availability of efficient and versatile
simulation tools.  In the last decade, the Relativistic Lattice Boltzmann method
(RLBM) has gained considerable interest in this context. To date, RLBM has been
derived and applied in the limit of vanishingly small Knudsen numbers $\rm Kn$,
defined as the ratio between the particles mean free path and a typical
macroscopic scale of the flow; available methods are increasingly inaccurate as
one increases the value of $\rm Kn$, moving towards beyond-hydrodynamic regimes.
On the other hand, beyond-hydro regimes are very relevant for QGP,  especially
with regard to their long-time evolution after the hydrodynamic epoch.
Furthermore, electron conduction in pure enough materials is almost ballistic,
and therefore more attuned to beyond-hydrodynamic descriptions.

The study these systems has been performed in the past as an eremitic expansion 
of the purely ballistic regime \cite{romatschke-epjc-2018, borghini-epjc-2018}. 
In this work we propose instead an extension of RLBM that builds on the hydrodynamic 
regime to further enhance its efficiency in the rarefied gas regime. 

The extension of RLBM to the study of rarefied gases has been previously
considered in the work by Ambru\c{s} and Blaga \cite{ambrus-prc-2018}. Based
on off-lattice product-based quadrature rules,  their model allow for an
accurate description of one-dimensional flows beyond hydrodynamic regimes.

In this work, we extend the RLBM in order to further enhance its efficiency in
the rarefied gas regime.

For simplicity, in this paper we consider gases of massless particles in a $(2+1)$ space
time, but the same methodologies can be extended to more general equations of
state, suitable for  fluids consisting of non-zero mass particles in three space
dimensions. 

This paper is organised as follows: in the first part of
Sec.~\ref{sec:model-description} we review the main  concepts of relativistic
kinetic theory, which are instrumental for the subsequent description of the
Relativistic  Lattice Boltzmann Method. In Sec.~\ref{sec:mom.space-disc}, we dig
deeper into the definition of the model, by  describing in more detail a
momentum space discretization procedure which enables the beyond-hydro
capabilities  of the scheme. Finally, in Sec.~\ref{sec:num-results}, we present
numerical evidence of the capabilities of the scheme, while Sec. ~\ref{sec:conclusions} 
presents our conclusions and prospects of further development.

%===================================================================================================
\section{Model Description}\label{sec:model-description}
%===================================================================================================

In this work we consider a two-dimensional gas of massless particles; we use 
a $(2+1)$ dimensional Minkowsky space-time, with metric signature $\eta^{\alpha\beta}=diag(+,-,-)$.
We adopt Einstein's summation convention over repeated indices; Greek indices denote $(2+1)$ 
space-time coordinates and Latin indices two-dimensional spatial coordinates. 
All physical quantities are expressed in natural units, $c = k_{\rm B} = 1$.

%===================================================================================================
\subsection{Relativistic Kinetic Theory}\label{subsec:relativistic-kinetic-theory}
%===================================================================================================

The relativistic Boltzmann equation, here taken in the relaxation time approximation (RTA)
\cite{anderson-witting-ph-1974b,anderson-witting-ph-1974a}, governs the time evolution of
the single particle distribution function $f(x^{\alpha}, p^{\alpha})$, depending on space-time coordinates 
$x^{\alpha}=(t, \mathbf{x})$ and momenta $p^{\alpha}=(p^{0}, \mathbf{p})$, with $\mathbf{x}, \mathbf{p} \in \mathbb{R}^{2}$:
\begin{align}\label{eq:boltz_eq}
  p^{\alpha}\partial_\alpha f = - \frac{U^\alpha p_\alpha}{\tau} \left( f - f^{\rm eq} \right) \quad ;
\end{align}
$U^\alpha$ is the macroscopic fluid velocity, $\tau$ is the (proper)-relaxation time
and $f^{\rm eq}$ is the equilibrium distribution function, which we write in a general form as
\begin{align}\label{eq:feq_dist}
  f^{\rm eq} \propto \frac{1}{ z^{-1} exp{\left(\frac{U_\alpha p^\alpha}{T}\right) + \varepsilon } } , \quad z = exp{\left( \frac{\mu}{T} \right)} , 
\end{align}
with $T$ the temperature, $\mu$ the chemical potential, and $\varepsilon$ distinguishing between the Maxwell-J{\"u}ttner ($\varepsilon = 0$),
Fermi-Dirac ($\varepsilon = 1$) and Bose-Einstein ($\varepsilon = -1$) distributions. 
In what follows we will restrict ourselves to the Maxwell-J{\"u}ttner statistic; however we remark that the quadrature rules
introduced in the coming section are general and apply to Fermi or Bose statistics as well.  

The particle flow $N^\alpha$ and the energy-momentum tensor $T^{\alpha\beta}$, 
respectively the first and second order moment of the distribution function
\begin{align}\label{eq:moments}
  N^\alpha = \int f p^\alpha         \frac{\diff^2 p}{p_0} \quad , \quad\quad
  T^{\alpha\beta} = \int f p^\alpha p^\beta \frac{\diff^2 p}{p_0} \quad ;    
\end{align}
can be put in direct relation with a hydrodynamic description of the system.
The RTA in Eq.~\ref{eq:boltz_eq} is in fact compatible with the Landau-Lifshitz \cite{landau-book-1987} decomposition: 
\begin{align}\label{eq:ll-decomp}
  N^\alpha        &= n U^{\alpha} - \frac{n}{P+\epsilon} q^\alpha                                \quad , \\
  T^{\alpha\beta} &= (\epsilon+P)U^\alpha U^\beta - P \eta^{\alpha\beta} + \pi^{<\alpha\beta>}   \quad .
\end{align}
$n$ is the particle number density, $P$ the pressure field, $\epsilon$ the energy density, 
$q^{\alpha}$ the heat flux, and $\pi^{<\alpha\beta>}$ the pressure deviator. 

%===================================================================================================
\subsection{Relativistic Lattice Boltzmann Method}\label{subsec:RLBM}
%===================================================================================================

In this section, we briefly summarise the derivation of the relativistic Lattice Boltzmann method,
referring the reader to a recent review  \cite{gabbana-pr-2020} for full details.

The starting point in the development of the scheme is a polynomial expansion 
of the equilibrium distribution function (Eq.~\ref{eq:feq_dist}) :
\begin{equation}\label{eq:mj-feq-expansion}
  f^{\rm eq}(p^{\mu}, U^{\mu}, T) 
  = 
  \omega( p^0) \sum_{k = 0}^{\infty} a^{(k)}_{i_1\dots i_k}( U^{\mu}, T) J^{(k)}_{i_1\dots i_k} ( p^{\mu} )  \quad ,
\end{equation}
with $\{ J^{(k)}_{i_1\dots i_k}, k = 1,2,\dots \} $ a suitable set of polynomials, orthogonal with respect 
to the weighting function $\omega(p^0)$, and the expansion coefficients are:
\begin{equation}\label{eq:mj-projection-coefficients}
  a^{(k)}_{i_1\dots i_k}( U^{\mu}, T) 
  = 
  \int f^{\rm eq}( p^{\mu}, U^{\mu}, T)  J_{i_1\dots i_k}^{(k)}( p^{\mu} ) \frac{\diff^2 p}{p^0} \quad .
\end{equation}
The choice of the polynomials to be used in Eq.~\ref{eq:mj-feq-expansion}
is directly related to the specific form of the equilibrium distribution function taken into consideration.
A convenient choice for the weighting function $\omega(p^0)$ is given by
the equilibrium distribution in the fluid rest frame; this choice delivers 
the nice and desirable property that the first $N$ coefficients of the truncated version
of Eq.~\ref{eq:mj-feq-expansion} coincide with the first $N$ moments of the distribution.

The next step consists of defining a Gaussian-type quadrature, able to recover
exactly all the moments of the distribution up to a desired order $N$.
The definition of the quadrature
is a crucial aspect, which will be covered in detail in the next section. For the moment
we assume that we can define a set $\{(w_i, p_i^{\mu}), i = 1,2, \dots\}  $ of weights and quadrature
nodes, allowing to formulate the discrete version of the equilibrium distribution function:
\begin{equation}\label{eq:mj-feq-expansion-truncated}
  f^{\rm eq}_i = f^{\rm eq}(p^{\mu}_i, U^{\mu}, T) 
  = 
  w_i \sum_{k = 0}^{N} a_{i_1\dots i_k}^{(k)}( U^{\mu}, T) J_{i_1\dots i_k}^{(k)}( p^{\mu}_i )  \quad .
\end{equation}

Consequently, Eq.~\ref{eq:boltz_eq} is discretized in a set of differential equations for the distributions $f_i$:
\begin{align}
  \partial_t f_i + \bm{v}^{i} \cdot \nabla f_i = - \frac{p_i^{\alpha} U_{\alpha}}{p^0_i \tau} \left( f_i - f_i^{\rm eq} \right) \quad ,
\end{align}
where $\bm{v}^{i} = \bm{p}^i / p_0^i $.
Next, by employing an upwind Euler discretization in time with step $\Delta t$ we derive the relativistic Lattice Boltzmann equation:
\begin{equation}\label{eq:discrete-rbe}
  f_i(\bm{x} + \bm{v}^{i} \Delta t, t + \Delta t) 
  = 
  f_i(\bm{x}, t) + \Delta t~ \frac{p_i^{\alpha} U_{\alpha}}{p^0_i \tau} (f_i^{\rm eq} - f_i(\bm{x}, t) ) \quad .
\end{equation}

From an algorithmic point of view the time evolution of the above equation can be split into
two parts, respectively the \textit{streaming} step, in which information is propagated to the neighboring sites,
and the \textit{collision} step, in which the collisional operator is locally applied to each grid point.

More formally, the streaming step can be defined as
\begin{align}\label{eq:streaming}
  f_i^*(\bm{x}, t) = f_i(\bm{x} - \bm{v}^{i} \Delta t, t) \quad ,
\end{align}
where information is moved at a distance $\Delta x = \bm{v}^{i} \Delta t$, which in general might not
define a position on the Cartesian grid.

In such cases it is therefore necessary to implement an interpolation scheme.
In this work we adopt a simple bilinear interpolation scheme:
\begin{align}\label{eq:interpolation}
  f_i (\bm{x} - \bm{v}^{i} \Delta t, t) &= \frac{1}{\Delta x \Delta y} \Bigg\{       \notag \\
     & f_i(\bm{x}             ~-\bm{r_x}            ~-\bm{r_y}, t) 
     \Big( \phantom{~1 -} \Delta t \big| v^i_x \big| \Big) 
     \Big( \phantom{~1 -} \Delta t \big| v^i_y \big| \Big)      \notag \\
     & f_i(\bm{x}   \phantom{~-\bm{r_x}}            ~-\bm{r_y}, t) 
     \Big(           1 -  \Delta t \big| v^i_x \big| \Big) 
     \Big( \phantom{~1 -} \Delta t \big| v^i_y \big| \Big)      \notag \\
     & f_i(\bm{x}             ~-\bm{r_x}   \phantom{~-\bm{r_y}}, t) 
     \Big( \phantom{~1 -} \Delta t \big| v^i_x \big| \Big) 
     \Big(           1 -  \Delta t \big| v^i_y \big| \Big)      \notag \\
     & f_i(\bm{x}   \phantom{~-\bm{r_x}}   \phantom{~-\bm{r_y}}, t) 
     \Big(           1 -  \Delta t \big| v^i_x \big| \Big) 
     \Big(           1 -  \Delta t \big| v^i_y \big| \Big) \Bigg\} 
\end{align}
with 
\begin{align}
  \bm{r_x} = sgn(v^i_x) \Delta x \bm{\hat{x}}  \\
  \bm{r_y} = sgn(v^i_y) \Delta y \bm{\hat{y}}
\end{align}

Next, one needs to compute the macroscopic fields, starting from the moments of the particle distribution functions, which 
thanks to a Gaussian quadrature can be defined as discrete summations:
\begin{align}\label{eq:discrete_sum_moments}
  N^\alpha = \sum_i^{N_{\rm pop}} p^\alpha_i f_i      \quad , \quad  T^{\alpha\beta} = \sum_i^{N_{\rm pop}} p^\alpha_i p^\beta_i f_i \quad .
\end{align}
From the definition of the energy-momentum tensor in the Landau frame, we compute the energy density $\epsilon$
and the velocity vector $U^{\alpha}$, by solving the eigenvalue problem: 
\begin{equation}
  \epsilon U^{\alpha} = T^{\alpha \beta} U_{\beta} \quad .
\end{equation}
The particle density $n$ can be then calculated using the definition of the first order moment, while pressure and 
temperature are obtained via a suitable equation of state; in this work we use the ideal equation of state (consistent with the J\"uttner distribution):
\begin{equation}
  \epsilon = 2 P = 2 n T \quad .
\end{equation}
The macroscopic fields allow in turn to compute the local equilibrium distribution and finally one applies the relaxation time collisional operator:
\begin{align}\label{eq:collision}
  f_i(\bm{x}, t + \Delta t) = f_i^*(\bm{x}, t) + \Delta t~ \frac{p_i^{\alpha} U_{\alpha}}{p^0_i \tau} (f_i^*(\bm{x}, t) - f_i^{\rm eq}) \quad .
\end{align}

%===================================================================================================
\subsection{ Momentum space discretization }\label{sec:mom.space-disc}
%===================================================================================================

As discussed in the previous section, the definition of a Gaussian-type
quadrature represents the cornerstone in the definition of a Lattice Boltzmann
method, since it allows the {\it exact}  calculation of integrals of the form of
Eq.~\ref{eq:mj-projection-coefficients} as discrete sums over the  discrete
nodes of the quadrature.
In the framework of RLBM, one distinguishes between two approaches in the
definition of  quadrature rules, each with advantages and
disadvantages: 
 i) On-lattice Lebedev-type quadrature rules 
ii) Off-lattice product-based quadrature rules.

On-lattice quadrature rules \cite{mendoza-prd-2013,gabbana-pre-2017,gabbana-cf-2018}
allow retaining one of the main LBM features, namely perfect streaming. Indeed, by requiring 
that all quadrature points lie on a Cartesian grid, it follows directly that at each time step
information is propagated from one grid cell to a neighbouring one, with two desirable side-effects:
i) super luminal propagation is ruled out by construction, and  ii) no artificial 
dissipative effects emerge, since there is no need of interpolation.

On the other hand, off-lattice quadratures, typically developed by means of product rules of Gauss-Legendre 
and/or Gauss-Laguerre quadratures  \cite{romatschke-prc-2011,coelho-cf-2018,ambrus-prc-2018}, offer
the possibility of handling more complex equilibrium distribution functions and to extend the 
applicability of the method to regimes that go beyond the hydrodynamic one. 
Conversely, the price to pay when going off-lattice is the requirement of an interpolation scheme. 
This makes it so that the advantages of on-lattices schemes represent the price to pay when going off-lattice. 

\begin{figure}[tbh]
  \centering
  \includegraphics[width=0.99\textwidth]{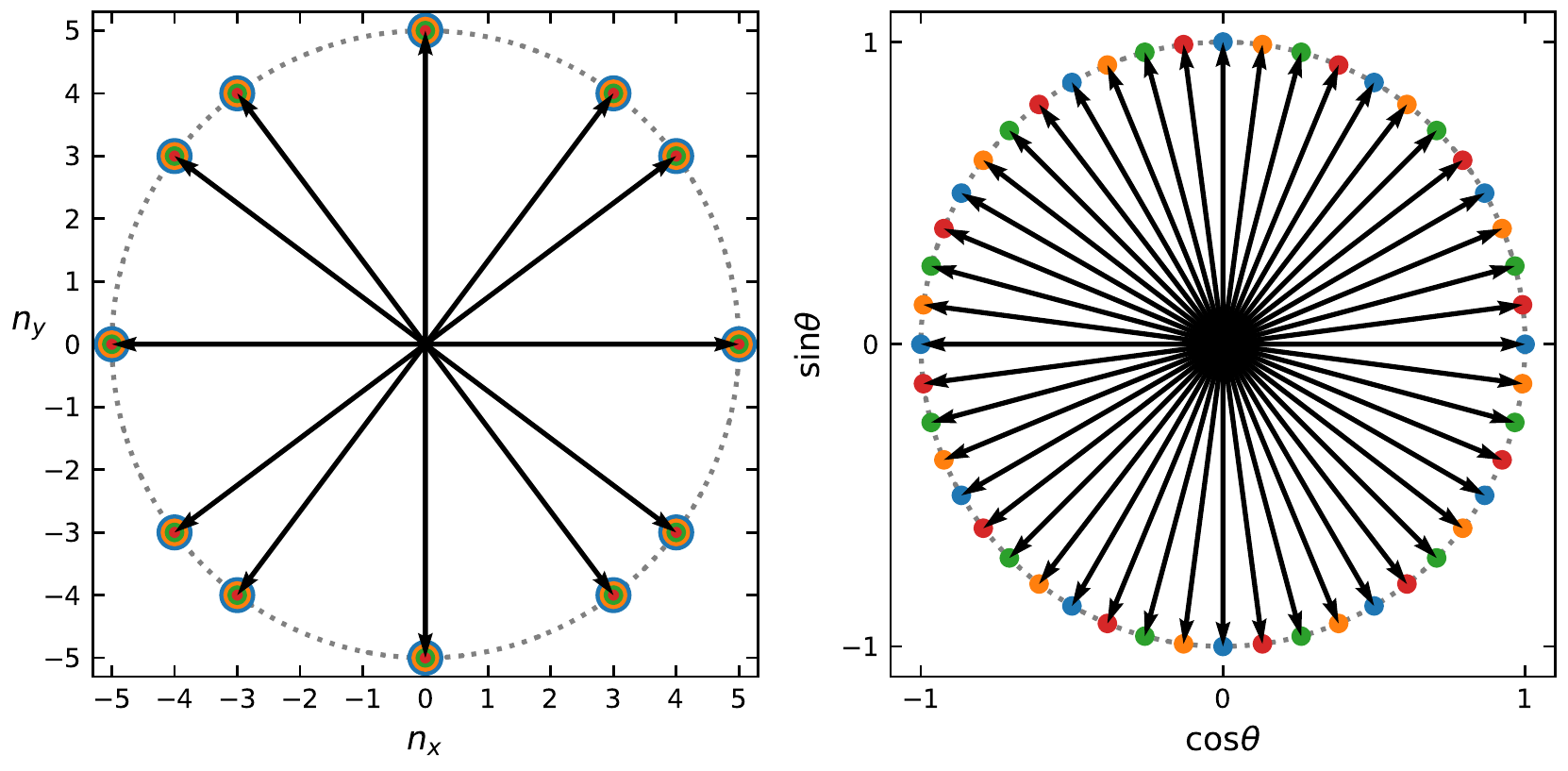}
  \caption{\small Two examples of stencils compatible with a third order
     quadrature. The arrows represent the discrete velocities
     $\vec{n}_i$,  while the different
     colors stand for different energy values $p^0_j$.  For an on-lattice quadrature (left panel)
     the velocity vectors of all energy shells lie
     at the intersection between the Cartesian grid and a circle of radius
     $5$. In an off-lattice example (right panel) the
     different energy shells are displaced in such a way that the vectors
     forming the stencil span uniformly the unit circle. In both cases,
     the total number of discrete components is $N_{\rm pop}=48$.
          } 
  \label{fig:1}
\end{figure}

For the definition of on-lattice quadratures, one can follow the so called method of quadrature
with prescribed-abscissas \cite{philippi-pre-2006}. 
In practice, one needs to find the weights and the abscissae of a quadrature able 
to satisfy the orthonormal conditions, up to the desired order:
\begin{align}\label{eq:mj-orthogonal-conditions}
  \int \omega(p^0) J_{i_1\dots i_m}^{(m)}( p^{\mu} ) J_{j_1\dots j_n}^{(n)}( p^{\mu} ) \frac{\diff^2 p}{p^0} 
  &=
  \sum_{i=1}^{N_{\rm pop}} w_i J_{i_1\dots i_m}^{(m)}( p^{\mu}_{i} )J_{j_1\dots j_n}^{(n)}( p^{\mu}_{i} ) \nonumber\\
  &= 
  \delta_{mn} \delta_{i_1 j_1} \dots \delta_{i_n j_m}  \quad ;
\end{align}
where $p^{\mu}_{i}$ are the discrete momentum vectors. A convenient parametrization of 
the discrete momentum vectors in the ultra-relativistic limit writes as follows:
\begin{equation}\label{eq:}
  p^{\mu}_{i,j} = p^0_j \left(1,  \frac{\vec{n}_i}{|| \vec{n}_i ||} \right), 
\end{equation}
where $\vec{n}_i \in \mathcal{Z}^2$ are the vectors forming the stencil, which are to be found at the intersection
between the Cartesian grid and a circle (or a sphere in (3+1) dimensions). 
Massless
particles travel at the speed of light irrespective of their energy, so this set of vectors can be assigned to different energy shells, 
each labeled via the index $j$, which are properly chosen in such a 
way that Eq.~\ref{eq:mj-orthogonal-conditions} returns valid solutions for the weights $w_i$. 
Note that vectors $\vec{n}_i$ must all have the same length $|| \vec{n}_i ||$, so information is correctly propagated at the speed of light. 
Finally, an $N$-order quadrature rule needs a minimum of $N + 1$ energy shells, to recover the moments exactly. 
Therefore, following the procedures adopted in \cite{gabbana-pr-2020,mendoza-prd-2013}, we 
select such shells as the zeros of the orthogonal polynomial $J_{0 \dots 0}^{(N+1)}(p^0)$. 

In order to define a quadrature rule on a Cartesian grid it is expedient to use
the same set of velocity vectors for the different energy shells; this allows to
achieve enough degrees of freedom such to  satisfy the orthonormal conditions in
Eq.~\ref{eq:mj-orthogonal-conditions} by using vectors of a relatively small
lenght.
In the left panel of Fig.~\ref{fig:1} we show an example of a quadrature recovering up to the third order moments 
of the distribution function, using vectors of lenght $5$. Extending the procedure to higher orders 
leads to stencils unviable for practical computational purposes, since already going to the fourth order would require
using vectors of lenght $5 \sqrt{13}$.

\begin{figure}[h!]
  \centering
  \includegraphics[width=0.99\textwidth]{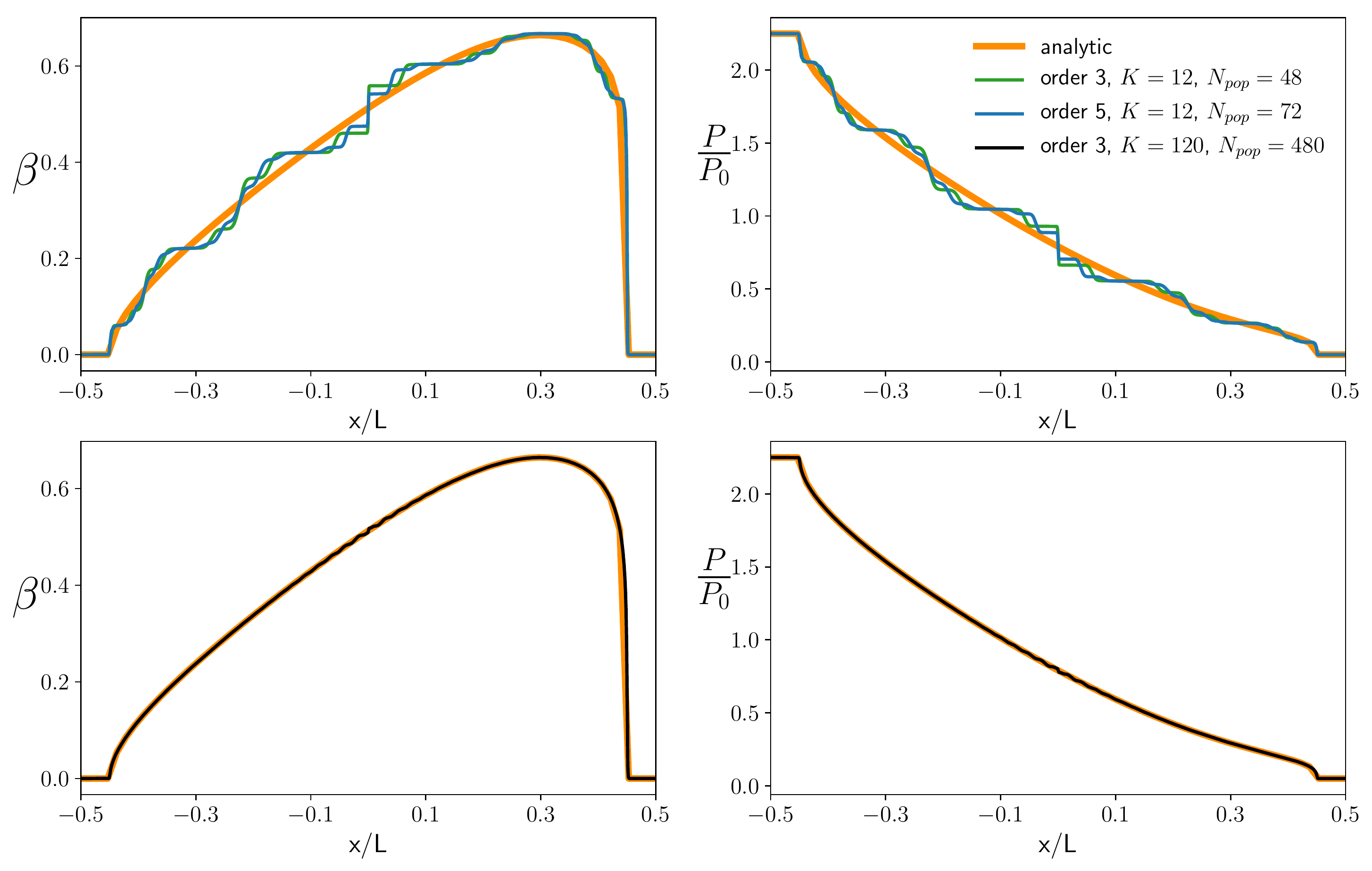}
  \caption{\small Mono-dimensional Sod Shock tube benchmark in the free-streaming regime 
           ($\rm{Kn} \to + \infty $, reached by setting $\tau \to +\infty$ in Eq.~\ref{eq:discrete-rbe}) 
           at time $t/t_{max} = 0.9$ with grid size $2000 \times 1$. 
           The analytic velocity and pressure fields given in \ref{appendix-1} (orange line) 
           are confronted with different numerical results produced using 
           different off-lattice stencils. In the top panels, the analytic solution is confronted with 
           a third order stencil with $K = 12$ (green line) and a fifth order quadrature with the same value 
           of $\rm K$ (blue line). These panels give no evidence of an increase in the quality of the solution 
           when increasing the order of the quadrature. In the bottom panels, the analytic solution is confronted 
           with a third order stencil with $K=120$, that accurately reproduces analytic results. 
  } 
  \label{fig:2}
\end{figure}

It is then clear that in order to recover the higher orders of the distribution it is necessary
to relax the condition of on-lattice streaming. Furthermore, when moving off-lattice it becomes 
convenient to assign different subsets to the different energy shells. For example, the stencil in the right panel 
of Fig.~\ref{fig:2} allows the definition of a quadrature rule that has the same number of discrete components, and
the same accuracy order of its on-lattice counterpart, but with a higher level of isotropy.

In order to define these off-lattice quadratures, one starts from
the observation that the orthonormal conditions in
Eq.~\ref{eq:mj-orthogonal-conditions} are equivalent to requiring the exact calculation of integrals of the form
\begin{align}\label{eq:quad-integrals}
  I^{\alpha_1\dots\alpha_k} = \int \omega(p^0) p^{\alpha_1} \dots p^{\alpha_k} \frac{\diff^2 p}{p^0} \quad .
\end{align}
for all $k \leq 2 N$. 

For an ultra-relativistic gas one has $p^0 = |\bm{p}|$, so it 
is useful to adopt polar coordinates and break down 
the integrals of Eq.~\ref{eq:quad-integrals} into a radial part and an angular one:
\begin{align}\label{eq:integrals-pol-coord-ultra}
  I^{\alpha_1 \dots \alpha_k} \propto 
  \left( \int_0^{\infty} e^{-\frac{p}{T}} p^{k} dp                       \right) 
  \left( \int_0^{ 2 \pi} (\cos \theta)^{k_1} (\sin \theta)^{k_2} d\theta \right) \quad .
\end{align}
with $ 0 \leq k_1 + k_2 \leq k $. 
We form the quadrature rule as a product rule: the Gauss-Laguerre rule is the most natural choice for
the radial component of Eq.~\ref{eq:integrals-pol-coord-ultra}, while for the angular part
we consider a simple mid-point rule (since the angular integral can be reworked using basic 
trigonometry as a sum of integrals of circular functions of maximum degree $2N$):
\begin{align}
  p_{ij}^{\mu} = 
  \begin{pmatrix}
    p_i               \\
    p_i \cos \theta_j \\ 
    p_i \sin \theta_j 
  \end{pmatrix} \quad ,
  &&
  w_{ij} = w^{(p)}_i w^{(\theta)}_j \quad ,
  &&
  \forall 0 \leq i \leq N, 0 \leq j \leq 2N \quad .
\end{align}
where $\{p_i, i = 1, 2, \dots N + 1\}$ are the roots of $L_{N+1}(p)$, the Laguerre polynomial of order $N+1$, and 
\begin{align}
  \theta_j        &= j \frac{2\pi}{2N+1}                              \quad , \\ 
  w^{(\theta)}_j  &= \frac{2 \pi}{2N + 1}                             \quad , \\
  w^{(p)}_i       &= \frac{p_i}{(N+2)^2 [L_{N+2}(p_i)]^2}             \quad .
\end{align}
The total number of points in the quadrature is $N_{\rm pop} = (N+1)(2N+1)$. 

In order to move to high Knudsen numbers this level of discretisation is however
not sufficient  to properly describe the dynamics of the system. Larger and more
evenly distributed sets of discrete velocities are needed, in order to cover
the velocity space in a more uniform way.

A possible solution is to increase the order of the angular quadrature, i.e.
raise the number of velocities per energy shell, even if this comes at  an
increased computational cost. Additionally, a further move that seems to be
beneficial in increasing the quality of the solution  without effectively
increasing the number of discrete velocities is the decoupling of radial and
angular abscissae. 

In fact, once the required quadrature orders needed to recover the requested hydrodynamic moments are met, the restriction 
of using the same sub-stencils $\theta_j$ for every energy shell $p_i$ can be lifted, and the isotropy 
of the model can be enhanced with no need of increasing the overall quadrature order. 

In $(2+1)$ dimensions this is easily achieved by rotating the sub-stencils related to different energy shells, in
such a way that the discrete velocities cover the velocity space in the most homogeneous possible way. 

With these two recipes in mind, our quadrature becomes: 
\begin{align}\label{eq:off-lattice-quad}
  p_{ij}^{\mu} = 
  \begin{pmatrix}
    p_i                   \\
    p_i \cos \theta_{ij}  \\ 
    p_i \sin \theta_{ij} 
  \end{pmatrix} \quad ,
  &&
  w_{ij} = w^{(p)}_i w^{(\theta)}_j \quad ,
  &&
  \forall 0 \leq i \leq N, 0 \leq j \leq K - 1\quad .
\end{align}
where $\rm K$ can be chosen freely as long as $K \geq 2N+1$ and
\begin{align}
  \theta_{ij}     &= \left(j + \frac{i}{N+1} \right) \frac{2\pi}{K}    \quad , \\ 
  w^{(\theta)}_j  &= \frac{2 \pi}{K}                                   \quad , \\
  w^{(p)}_i       &= \frac{p_i}{(N+2)^2 [L_{N+2}(p_i)]^2}              \quad .
\end{align}

All together, there are $N_{\rm pop}=K(N+1)$ points. 
In the right panel of Fig.~\ref{fig:1} we compare an example of a quadrature obtained 
with this new method (right panel) with a more traditional on-lattice one.

%===================================================================================================
\section{Numerical results}\label{sec:num-results}
%===================================================================================================

%===================================================================================================
\subsection{Mono-dimensional Shock Waves}\label{subsec:mono-sod}
%===================================================================================================

We test the ability of our new numerical scheme to simulate beyond-hydrodynamic regimes, considering 
as a first benchmark the Sod shock tube problem, which has an analytic solution in the free streaming regime, 
derived in \ref{appendix-1}. 

In our numerical simulations we consider a tube defined on a grid of $L \times 1$ points. The tube is filled 
with a fluid at rest, and there is a discontinuity in the values of the thermodynamic quantities in the middle 
of the domain (that is, considering a $[-L/2,L/2]$ domain, at the value $x = 0$); 

By normalizing all quantities to appropriate reference values, we take

\begin{align}\label{eq:initial-cond-macro}
\left(~ \frac{P}{P_0},~ \frac{n}{n_0},~ \frac{T}{T_0},~ \beta~ \right) = 
\begin{cases}
  (2.25,~ 1.5,~ 1.5,~ 0.0)   \quad\quad x < 0 \\
  (0.05,~ 0.1,~ 0.5,~ 0.0)   \quad\quad x > 0
\end{cases}
\end{align}

Once the division between the two domains is removed, pressure and temperature differences  
develop into a mono-dimensional dynamics of shock - rarefaction waves traveling along the tube. 

\begin{figure}[h!]
  \centering
  \includegraphics[width=0.99\textwidth]{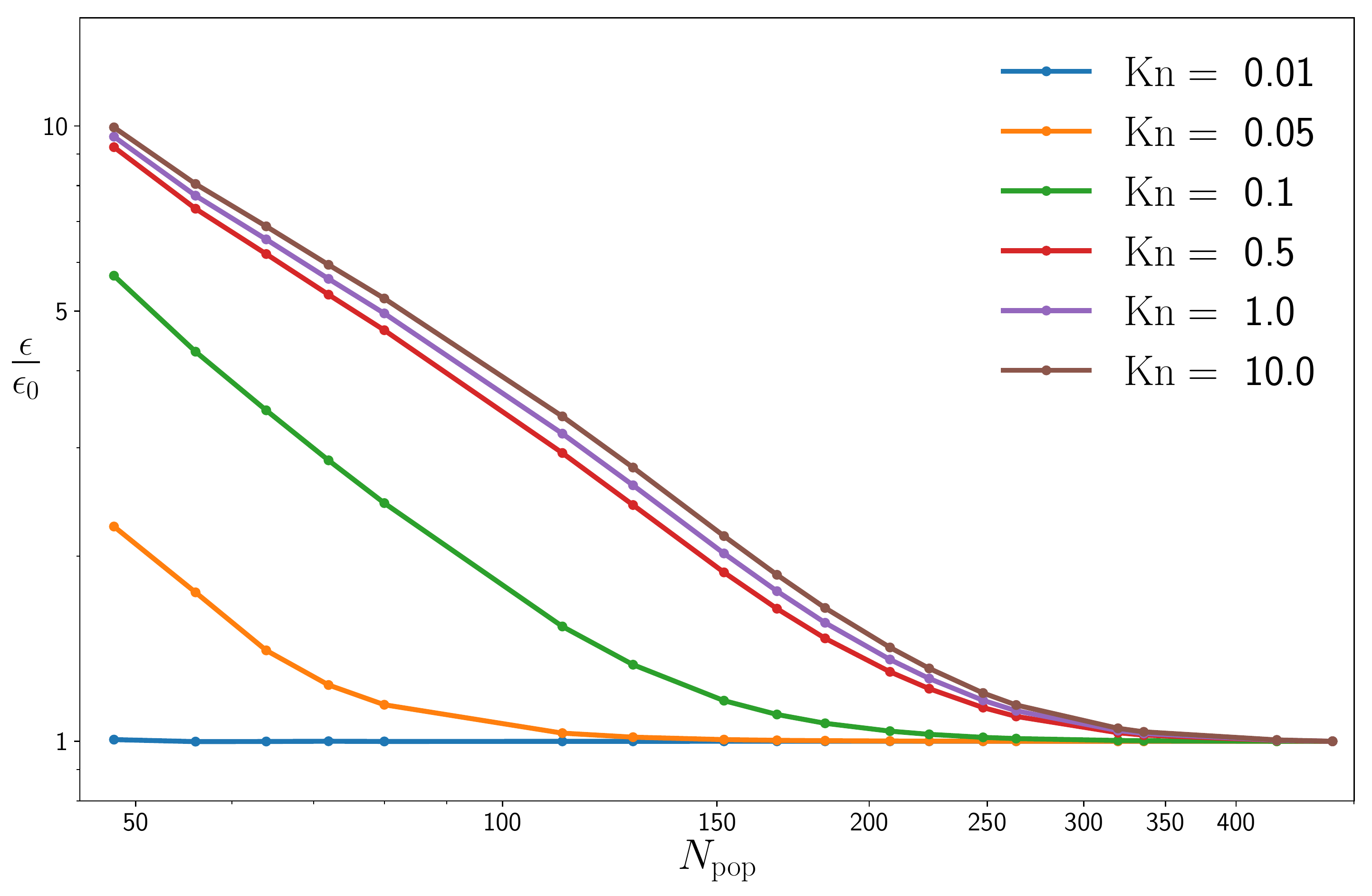}
  \caption{\small Comparison of the L2 difference between stencils with
     different values of $N_{\rm pop}$ on a grid size $2000 \times 1$  and
     a very high resolution instance with $N_{\rm pop} = 2400$ and grid
     size $4000 \times 1$ for different values of $\rm Kn$. For values of
     $\rm Kn$ in the hydrodynamic regime, the quality of the solution does
     not depend significantly on the number of populations $N_{\rm pop}$
     and $\epsilon$ only depends   on spatial resolution.  As $\rm Kn$
     increases, $\epsilon$  depends on $N_{\rm pop}$ until the saturation point is
     reached, and the residual error depends again on spatial resolution. 
     The saturation point grows with $\rm Kn$, from $\sim 100$ for
     $\rm{Kn}=0.05$ to $\sim 350$ for $\rm{Kn} \sim 10$.  All $\epsilon$
     values are normalized with respect their asymptotic values
     $\epsilon_0$, which is of order $10^{-3}$.
  } 
  \label{fig:3}
\end{figure}

Fig. \ref{fig:2} shows a subset of the results of our simulation at time $t/t_{max} = 0.9$ 
($t_{max}$ being the time needed by the shock to reach the edge of the box),
for two different quadrature orders and for several choices of $\rm K$.  As
higher order quadratures naturally imply larger values of $\rm K$ it is in principle
debatable which is the main actor  leading to accurate results in the beyond
hydrodynamics regime.  Fig.~\ref{fig:2} provides an answer to this question. 
We first show that, for a comparable (and low) value of $\rm K$, quadratures of
different order lead to similar (and unsatisfactory) results; on the other hand,
limiting the quadrature order to $3$ but substantially increasing $\rm K$ (and
consequently $N_{pop})$ we obtain results in very good agreement with the
analytic solution. This provides strong evidence of the important result that the proper representation of these kinetic
regimes can be achieved by employing sufficiently dense velocity sets, even with
quadratures recovering only the minimum number of  moments of the particle
distributions able to provide a correct representation of the thermodynamical
evolution of the system.

Starting from the initial conditions defined in Eq.~\ref{eq:initial-cond-macro} we now extend our analysis to intermediate regimes 
characterized by finite values of the Knudsen number, with the aim of establishing a relation between $\rm{Kn}$ and the optimal 
choice for $\rm K$. In our simulations, we use a fixed value of the relaxation time $\tau$, and assume the following expression 
for the Knudsen number 
\begin{equation}
  \rm{Kn} = \frac{c ~ \tau}{L} \quad 
\end{equation}
the value for $\tau$ is properly rescaled when one wants to compare simulations with different $\rm L$ but equal $\rm Kn$.
We use quadrature rules of order $N=3$, with different values of $\rm K$, and compare
against a reference solution obtained solving the RTA with a highly refined discretization 
both in terms of momentum space and grid resolution. For the reference solution we use 
$K = 600$ and a grid size $4000 \times 1$. 
In order to quantify the accuracy of the result we introduce the parameter $\epsilon$,
the relative error computed in L2-norm of the macroscopic velocity profile
\begin{align}\label{eq:rel_err}
 \epsilon = \frac{|| \beta - \beta_{\rm ref} ||_2}{|| \beta_{\rm ref} ||_2}     \quad .
\end{align}
 
As expected, at low $\rm Kn$ values, $\epsilon$ stays constant as one increases the value of $\rm K$, and the differences between the two solutions are only 
due to the finer spatial resolution of $\beta_{\rm ref}$. When transitioning to beyond-hydro regimes, $\epsilon$ starts to exhibit a power 
law dependency with respect to $\rm K$ (and therefore $N_{\rm pop}$) since the low order momentum space discretization comes into play. 

The bottom line of this power law decay occurs once the size of the artifacts in the macroscopic profiles 
(i.e. the ``staircase'' effect visible in Fig.~\ref{fig:2}), become comparable with the grid spacing.
From that point on, $\epsilon$ stays constant as the spatial resolution error becomes preponderant over the velocity resolution one.  

As expected, the optimal choice for $\rm K$ grows as one transitions from the hydrodynamic to the ballistic regime. 
In any case, from Fig.~\ref{fig:3} it is possible to appreciate that the minimum number of $N_{\rm pop}$ never exceeds $\sim 350$ populations.
        
\begin{figure}[h!]
  \centering
  \includegraphics[width=0.99\textwidth]{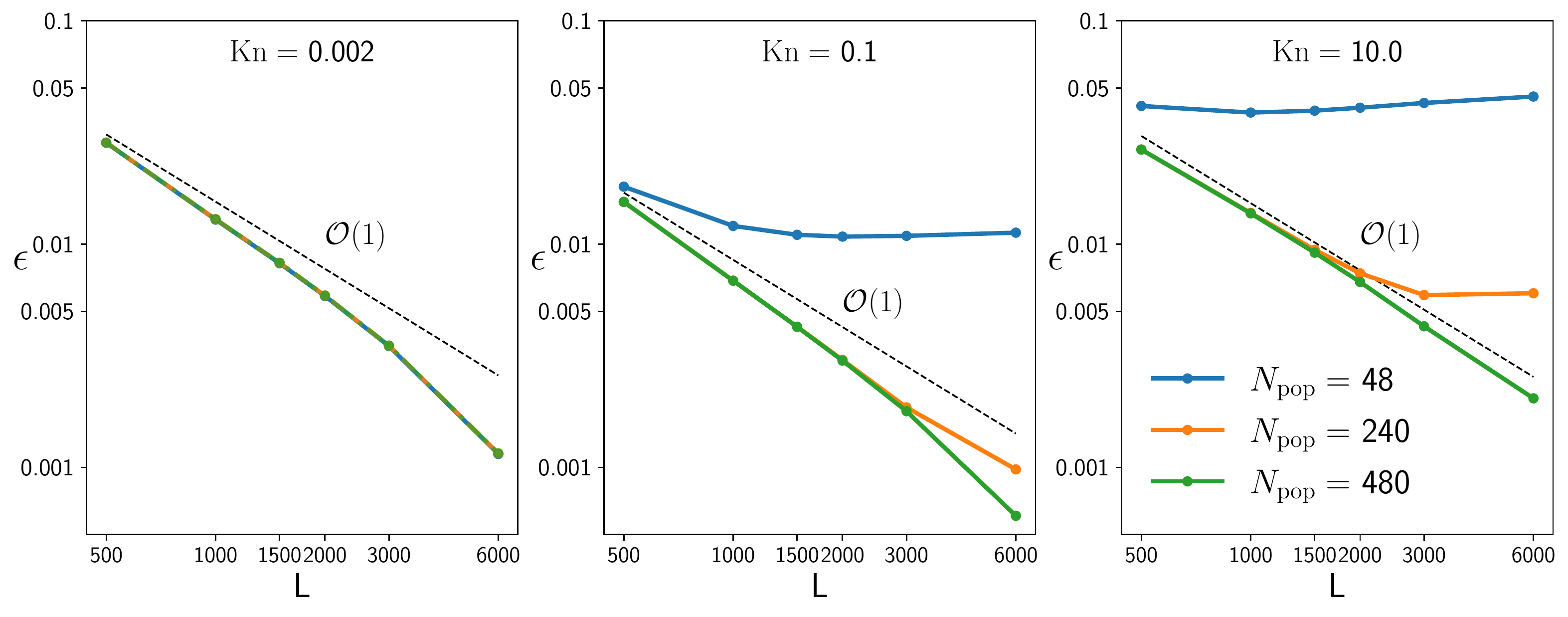}
  \caption{\small Convergence analysis of the numerical scheme in three
     different kinetic regimes, using stencils with $N_{\rm pop}$ $48$, $240$,
     and $480$. The comparison is performed with  respect to a high resolution
     simulation using $N_{\rm pop} = 2400$ (grid size $12000 \times 1$).  
     While in the hydrodynamic regime ($\rm{Kn} = 0.002$)
     the results do not depend on the number of discrete components employed,
     the dependence of the relative error $\epsilon$ on $N_{\rm pop}$ becomes
     evident as we transition towards the ballistic regime. The figures also clearly highlight that, 
     for the larger $\rm Kn$ values, the error almost reaches a ($N_{\rm pop}$ dependent) 
     plateau as the grid size becomes finer and finer.
  }  
  \label{fig:4}
\end{figure} 

We conclude this section with a convergence analysis of the method.
In Fig.~\ref{fig:4} we analyze the scaling of the relative error Eq.~\ref{eq:rel_err} as a function of 
the grid size in three different kinetic regime, comparing stencils with $N_{\rm pop}$ $48$, $240$, and $480$.
While in the hydrodynamic regime ($\rm{Kn} = 0.002$) the results do not depend on the number
of discrete components employed, the dependence of the relative error $\epsilon$ on $N_{\rm pop}$
becomes evident as we transition towards the ballistic regime. 
In this case, the constraint of moving particles along a finite number of directions introduces
a systematic error, which becomes dominant as one increases the grid resolutions. As a result,
when the momentum space is not adequately discretized the error stops scaling, eventually reaching almost a plateau value.
%     

%===================================================================================================
\subsection{Two-dimensional Shock Waves}\label{subsec:bi-sod}
%===================================================================================================

Purely bi-dimensional shock waves are commonly used as validation benchmarks in relativistic and non-relativistic 
\cite{suzuki-mnras-2016,delzanna-aa-2003,chen-jcp-2017,marti-lrca-2015} CFD solvers, since they provide a useful 
test bench to evaluate dynamics in the presence of sharp gradients. 

\begin{figure}[h!]
  \centering
  \includegraphics[width=0.99\textwidth]{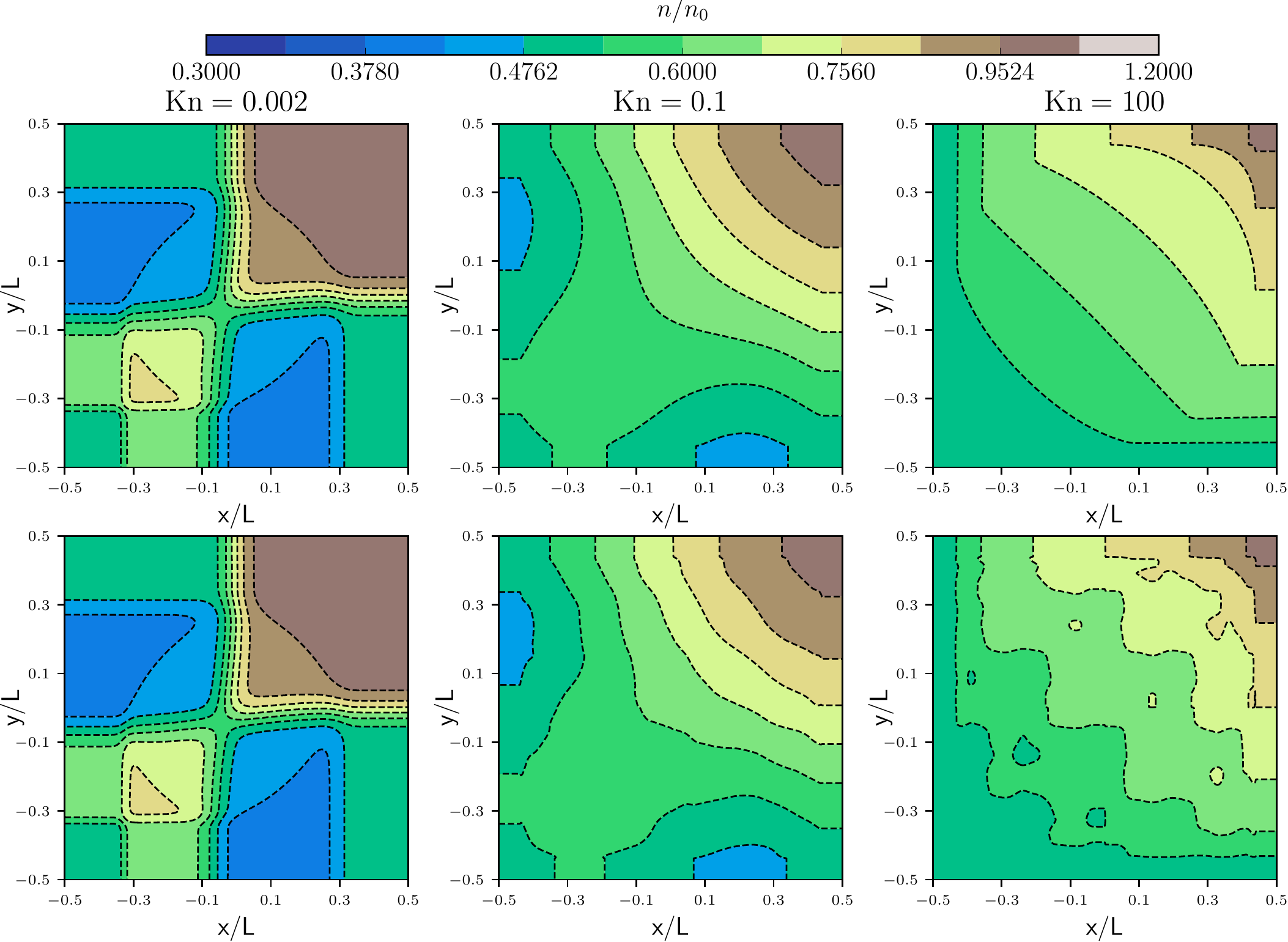}
  \caption{ \small Color plot of the particle density for the bi-dimensional Sod problem, in the presence of the initial conditions of 
           Eq.~\ref{eq:initial-cond-macro} and at time $t/t_{max} = 0.9$. The top panels show solutions for a spatial grid 
           with $1000 \times 1000$ lattice points, and $N_{\rm pop} = 2400$. The bottom panels have instead a spatial grid of 
           $500 \times 500$ lattice points, with $N_{\rm pop} = 48$. From left to right different kinematic regimes (different $\rm Kn$) 
           are explored. As $\rm Kn$ grows and the dynamic enters the beyond-hydrodynamic regime, $N_{\rm pop} = 48$ 
           performs poorly while sensible results are obtained with the large value of $N_{\rm pop}$. Contour lines are logarithmically spaced. 
           } 
  \label{fig:5}
\end{figure}

In a box domain of extension $[-L/2, L/2] \times [-L/2, L/2]$, we impose the initial conditions
\begin{align}\label{eq:initial-cond-macro}
\left(\frac{P}{P_0}, \frac{n}{n_0}, \frac{T}{T_0}, \beta_x, \beta_y\right) = 
\begin{cases}
  (0.5, ~~ 0.5, ~~ 1.0, ~~ 0.0, ~~ 0.0)   \quad, \quad x < 0 \quad y < 0   \quad, \\
  (1.0, ~~ 0.5, ~~ 2.0, ~~ 0.0, ~~ 0.1)   \quad, \quad x > 0 \quad y < 0   \quad, \\
  (1.0, ~~ 0.5, ~~ 2.0, ~~ 0.1, ~~ 0.0)   \quad, \quad x < 0 \quad y > 0   \quad, \\
  (1.0, ~~ 1.0, ~~ 1.0, ~~ 0.0, ~~ 0.0)   \quad, \quad x > 0 \quad y > 0   \quad. 
\end{cases}
\end{align}

Under these settings, the system develops into a square shock wave that travels toward the top-right part of the box. 
In Fig.~\ref{fig:5} we show a snapshot of the density field at time $t/t_{max} = 0.9$, 
for three different values of the Knudsen number, corresponding to a hydrodynamic regime ($\rm{Kn} = 0.002$), 
a transition regime ($\rm{Kn} = 0.1$), and an almost ballistic regime ($\rm{Kn} = 100$).
The top panels are obtained using a model employing a third order quadrature with $K = 600$, while the bottom one
uses $K = 12$. The two solvers are in excellent agreement when working in the hydrodynamic regimes, 
with artificial patterns emerging as we transition beyond hydro regimes for the case $K = 12$.
 
Similarly to what has been done in the previous section for the case of the mono-dimensional shock wave, 
we have investigated once again the dependency of the optimal choice for $\rm K$ with respect to $\rm{Kn}$.
This time the reference solutions have been calculated using a quadrature with $K=600$ and a grid of size 
$1000 \times 1000$. All other simulations employ a grid of size $250 \times 250$. 

\begin{figure}[h!]
  \centering
  \includegraphics[width=0.99\textwidth]{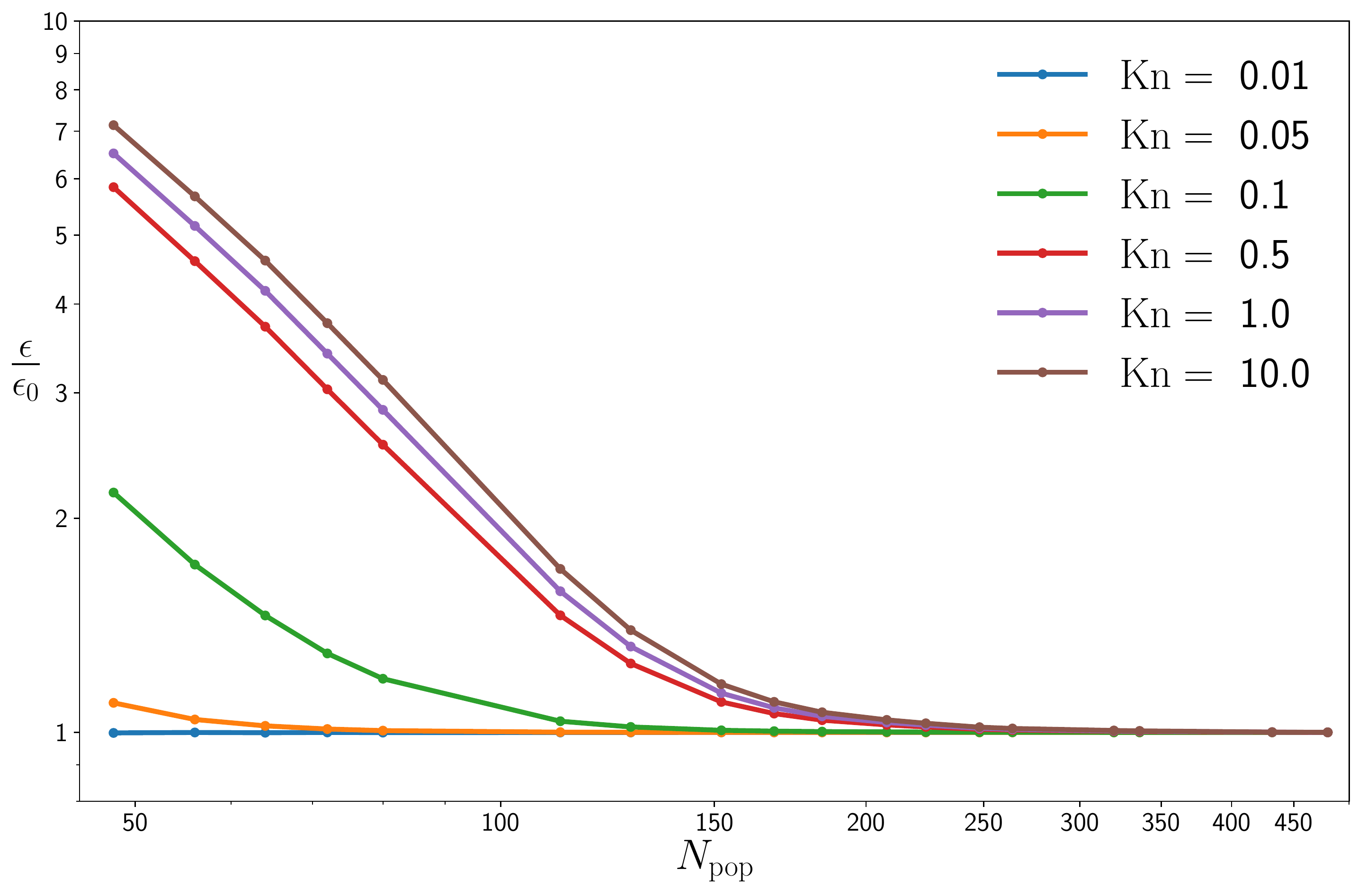}
    \caption{\small Comparison of the L2 difference between stencils at
      different values of $N_{\rm pop}$  (grid size $250 \times 250$) and a very
      high resolution instance with $N_{\rm pop} = 2400$ and  grid size $1000
      \times 1000$ for different values of $\rm Kn$. For values of $\rm Kn$ in
      the hydrodynamic regime, the quality of the solution does not depend
      significantly on $N_{\rm pop}$.   As $\rm Kn$ increases, $\epsilon$ starts
      to depend on  $N_{\rm pop}$ until the saturation point is reached. After this value, the
      stepping error is below  the grid resolution and can not be visualized
      anymore. The saturation value moves with $\rm Kn$, from $ <100$  in the
      case $\rm{Kn}=0.05$ to $\sim 250$ in the case $\rm{Kn} \sim 10$. 
      $\epsilon$ values are normalized with respect their asymptotic values
      $\epsilon_0$, which is of order $10^{-3}$. 
} 
  \label{fig:6}
\end{figure}

The results are presented in Fig.~\ref{fig:6}, and closely resemble those presented in the mono-dimensional case,
with the optimal choice for $\rm K$ happens to be consistent with that of a mono-dimensional flow,
and even at large values of $\rm{Kn}$ the minimum number of velocities to be taken in order to obtain a correct 
solution is in line with the figures obtained for the Sod tube.

%===================================================================================================
\section{Conclusion}\label{sec:conclusions}
%===================================================================================================

In this paper, we have presented a Relativistic Lattice Boltzmann Method for the simulation  
of gases of ultra-relativistic particles  in two spatial dimensions. 
The method is able to describe free-streaming dynamics ($\rm{Kn} \gg 1 $) as well as hydro-dynamics 
($\rm{Kn} \ll 1$). The simulation of beyond-hydro regimes is enabled by an off-lattice discretization technique of  
the momentum space, which comes at the price of introducing  some amount of numerical diffusivity. 

The procedure consists in adopting a product rule for the quadratures (strategy already adopted in the past, 
for example in \cite{ambrus-prc-2018}) and in the additional step of employing different velocity subsets 
for the different energy shells. 
In this way, a finer discretization of the two-dimensional velocity space is achieved, which is instrumental 
for simulations at high values of $\rm Kn$. 

The method has been benchmarked on two different realizations of the Sod shock tube problem, a 
popular benchmark in fluid dynamics. We have considered both mono and bi-dimensional flows,
also providing analytical solutions for the limiting ballistic case.

Our results show that it is possible to extend RLBM to beyond-hydro regimes, 
provided that a sufficient number of populations is used, independently of the quadrature order.
Also, an analysis on the minimum number of components of the stencils needed 
to provide accurate solutions has been conducted, arriving at the result that $N_{\rm pop} \sim 350$ 
is sufficient for the purpose of reproducing the correct dynamics in every regime. 

The numerical method developed in this paper is instrumental for the simulation of 
relativistic problems that transition toward beyond hydrodynamic regimes. 
Relevant examples in point are Quark Gluon Plasmas produced in heavy ion collisions, and 
electron transport in exotic materials such as graphene. 

Much is left for the future. 
To start, it would be important to evaluate the computational performance 
of the method against those of standard Monte Carlo approaches when working at finite Knudsen numbers.
Furthermore, a direct application of the method to the study
of beyond-hydro regimes in graphene will require the definition of appropriate boundary condition
schemes capable of reproducing experimental results \cite{guo-pnas-2017, kumar-np-2017, kiselev-prb-2019}.
Finally, the extension of the method to three spatial dimensions, 
as well as to gases of massive particles, will be reported in an extended version of the present paper.

%===================================================================================================
\section*{Acknowledgments}
%===================================================================================================
The authors would like to thank Luciano Rezzolla and Lukas Weih for useful discussions. 
DS has been supported by the European Union's Horizon 2020 research and
innovation programme under the Marie Sklodowska-Curie grant agreement No. 765048.
SS acknowledges funding from the European Research Council under the European
Union's Horizon 2020 framework programme (No. P/2014-2020)/ERC Grant Agreement No. 739964 (COPMAT).
AG would like to thank professor Michael G\"unther and professor Matthias Ehrhardt for their kind hospitality at Wuppertal University.
All numerical work has been performed on the COKA computing cluster at Universit\`a di Ferrara. 

%===================================================================================================
\section*{Data availability}
%===================================================================================================
The data underlying this article will be shared on reasonable request to 
the corresponding author.
%===================================================================================================

%===================================================================================================
\appendix
%===================================================================================================

%===================================================================================================
\section{ Analytic solution of the mono-dimensional Sod shock tube in the free streaming regime }
\label{appendix-1}
%===================================================================================================

We present here the analytic solution of the Sod shock tube problem in the free-streaming regime, for a gas of 
ultra-relativistic particles. The calculations closely follow the steps highlighted in \cite{ambrus-prc-2018}
for the three-dimensional case.
 
We consider the relativistic Boltzmann equation in the free streaming regime:
\begin{align}\label{eq:boltz-eq-free}
  p^\alpha \partial_\alpha f = p^0 \partial_t f + p_x \partial_x f = \partial_t f + v_x \partial_x f = 0 \quad ,
\end{align}
and the following initial conditions for the macroscopic fields 
\begin{align}\label{eq:initial-cond-macro}
(P, n, T, \beta) = 
\begin{cases}
  (P_{\rm L}, n_{\rm L}, T_{\rm L}, 0)   \quad\quad x < 0 \\
  (P_{\rm R}, n_{\rm R}, T_{\rm R}, 0)   \quad\quad x > 0
\end{cases}
\end{align}

By introducing the self-similar variable $w = x / t$ one can write the solution of Eq.~\ref{eq:boltz-eq-free} by
distinguishing two different regions, respectively the unperturbed one for $|w| > 1$, and the perturbed one 
at $|w| \leq 1$:
\begin{align}
  f(w, p, v_x) = 
  \begin{cases}
    f^{\rm eq}_{\rm L} \phantom{~+ \theta(w-v_x) (f^{\rm eq}_{\rm R} - f^{\rm eq}_{\rm L}) \quad\quad}   w < - 1   \\
    f^{\rm eq}_{\rm L}           + \theta(w-v_x) (f^{\rm eq}_{\rm R} - f^{\rm eq}_{\rm L}) \quad\quad   |w| \leq 1 \\
    f^{\rm eq}_{\rm R} \phantom{~+ \theta(w-v_x) (f^{\rm eq}_{\rm R} - f^{\rm eq}_{\rm L}) \quad\quad}   w > 1
  \end{cases}
\end{align}

In order to define the macroscopic profiles in the perturbed region, we need to calculate integrals
in the form of Eq.~\ref{eq:moments}. The full form for $N^\alpha$ and $T^{\alpha\beta}$ in the 
perturbed region is given by:
\begin{align}
  N^0 &= n_{\rm L}  + i_0(w) \frac{n_{\rm R}-n_{\rm L}}{2\pi}      \\
  N^x &=              i_1(w) \frac{n_{\rm R}-n_{\rm L}}{2\pi}      \\
  N^y &= 0                                          
\end{align}
\begin{align}
  T^{00} &= P_{\rm L}  + i_0(w) \frac{P_{\rm R}-P_{\rm L}}{\pi}            \\
  T^{0x} &=              i_1(w) \frac{P_{\rm R}-P_{\rm L}}{\pi}            \\
  T^{0y} &= 0                                                           \\
  T^{xx} &= P_{\rm L}  +      i_2(w)  \frac{P_{\rm R}-P_{\rm L}}{\pi}      \\
  T^{yy} &= P_{\rm L}  + (i_0(w)-i_2(w)) \frac{P_{\rm R}-P_{\rm L}}{\pi}      \\
  T^{xy} &= 0            
\end{align}

where 
\begin{align*}
  i_0(w) = &\int_0^{2\pi}                 \theta(w - \cos\alpha )   \diff \alpha = 2 \pi - 2 \arccos(w)               \\
  i_1(w) = &\int_0^{2\pi} (\cos   \alpha) \theta(w - \cos\alpha )   \diff \alpha =-2 \sqrt{1- w^2}                    \\
  i_2(w) = &\int_0^{2\pi} (\cos^2 \alpha) \theta(w - \cos\alpha )   \diff \alpha =\pi - w \sqrt{1- w^2} - \arccos(w)  \\
\end{align*}
From the above relations, the thermodynamic quantities can be obtained.

\biboptions{sort&compress}
\bibliographystyle{elsarticle-num}
\bibliography{biblio}

\end{document}